\documentclass[preprint,showpacs,showkeys,amsmath,amssymb]{revtex4-1}

\usepackage{xcolor}
\usepackage{setspace}
\usepackage{graphicx}
\usepackage{epsfig}
\usepackage[latin1]{inputenc}
\usepackage[active]{srcltx}
\usepackage{dcolumn}
\usepackage{bm}
\usepackage{epstopdf}

\begin{document}


\preprint{IST 4.2015-Pinheiro}

\title[DEF: The Physical Basis of Electromagnetic Propulsion]{DEF: The Physical Basis of Electromagnetic Propulsion}

\author{Mario J. Pinheiro}
\address{Department of Physics, Instituto Superior Tecnico - IST, Universidade de Lisboa - UL, Av. Rovisco Pais, \& 1049-001
Lisboa, Portugal
} \email{mpinheiro@tecnico.ulisboa.pt}




\date{\today}
\begin{abstract}
The very existence of the physical vacuum provides a framework to propose a general mechanism for propelling bodies through an agency of electromagnetic fields, that seat in that medium. When two sub-systems of a general closed device interact via nonlocal and retarded electromagnetic pulses, it is easily shown that they give a nonzero force, and that only tend to comply with the action-to-reaction force in the limit of instantaneous interactions. The arrangement of sub-systems provide a handy way to optimize the unbalanced EM force with the concept of impedance matching. The general properties of the differential electromagnetic force (DEF) are the following: i) it is proportional to the square of the intensity and to the angular wave frequency $\omega$; ii) to the space between the sub-systems (although in a non-linear manner); iii) it is inversely proportional to the speed of interaction; iv) when the two sub-systems are out-of-phase, DEF is null. The approach is of interest to practical engineering principles of propulsion since it offers guiding principles useful to build prototypes.
\end{abstract}
\maketitle

\section{Introduction}

It is clear the inadequacy of chemical-fuel propulsion engines and the pressing need for new energy sources to secure our societies. Tsiolkovsky's formula shows that the velocity attained by a rocket is the product of the exhaust velocity by the logarithm of the ratio of the launching mass to the mass of the payload. It indicates that the launch weight is exaggerated. New solutions may be supplied by plasma physics, or exotic arrangements of electric currents, in order to generate thrust forces based on local violations of the action-to-reaction law.

Recent investigations in out of equilibrium systems~\cite{Pinheiro_SR} have shown that the linear momentum should read:
\begin{equation}\label{eq01}
\mathbf{p}=m\mathbf{v}_e+q\mathbf{A}-mT\frac{\partial \overline{S}}{\partial \mathbf{p}}.
\end{equation}
Eq.~\ref{eq01} was obtained using a new theoretical technique based on maximization of entropy. In this framework, it was assumed that all particles have the same drift velocity and that they turn all with the same angular velocity , the center of mass of each particle moving with the same macroscopic velocity . The last term of Eq.1 represents the gradient of the entropy, in an out of equilibrium process, and  is the transformed function [1]. The last term is a new momentum term, physically understood as a kind of ``entropic momentum" since it is ultimately associated to the information exchanged with the medium on the physical system viewpoint (e.g., momentum that eventually is radiated by the charged particle). Lorentz's equations don't change when time is reversed, but when retarded potentials are applied the time delay of electromagnetic signals on different parts of the system do not allow perfect compensation of internal forces, introducing irreversibility into the system. Ref.[1] gives several examples of irreversible (out of equilibrium) phenomena that do not comply with action-reaction law. We can argue that the momentum is always a conserved quantity provided that we add the appropriate term, in order Newton's third
Law can be verified. This apparent "missing symmetry" might result because matter alone does not form a closed system and we need to include the physical vacuum in order to restore lost symmetry. So, when we have two systems   and   interacting via some kind of force field , the reaction from the vacuum must be included as a sort of bookkeeping device:
\begin{equation}\label{eq02}
\mathbf{F}_{12}^{matter}=-\mathbf{F}_{21}^{matter} + \mathbf{F}^{vacuum}.
\end{equation}
	
We may assume the existence of a physical vacuum (see, e.g., Ref.~\cite{Lee_1981}) probably well described by a spin-0 field $\phi(x)$ whose vacuum expectation value is not zero, vacuum $\sim \phi(x)$.

The entropic term in Eq.~\ref{eq01} can be clarified in the case of an electromagnetic system if we consider, not merely a point particle, but an extensive body, suffice to consider an expansion of the vector potential (now, in gaussian units) around a given point $\mathbf{r}'$:
\begin{equation}\label{eq03}
\mathbf{p}=m\mathbf{v}+\frac{1}{c} \int_V \rho_e(r') \mathbf{A}(\mathbf{r}+\mathbf{r}',t)d^3r',
\end{equation}
which gives after Taylor expansion
\begin{equation}\label{eq04}
\mathbf{p}(\mathbf{r},t)=m\mathbf{v}(\mathbf{r},t)+\frac{q}{c}\mathbf{A}(\mathbf{r},t)+ \nabla^2 \mathbf{A}(\mathbf{r},t) I \gamma,
\end{equation}
where $\gamma$ is the gyromagnetic ratio ($\gamma =q/mc$), that we rewrite for an extended particle under the form:
\begin{equation}\label{eq05}
\gamma = \frac{1}{cI}\int_V \rho_e(r')r^{'2} d^3 r'.
\end{equation}
Here, $I$ denotes the moment of inertia of the ``body". Therefore, we may identify the last term of Eq.~\ref{eq01} with the last term of Eq.~\ref{eq04}, interpreting the entropic momentum as due to an arrangement of a certain number of charged particles in motion inside a given volume $V$ and not necessarily with ``dissipation". Therefore
\begin{equation}\label{eq06}
mT \frac{\partial \overline{S}}{\partial \mathbf{p}} \equiv \gamma I \nabla^2 A.
\end{equation}
We can translate from gaussian to SI units by means of the replacement $\frac{1}{c} \to \frac{\mu}{4 \pi}$. Recall that
\begin{equation}\label{eq07}
\nabla^2 \mathbf{A}=- mu \mathbf{J}+\mu \epsilon\frac{\partial^2 \mathbf{A}}{\partial t^2}.
\end{equation}
It is a matter of practical engineering to develop the appropriate fields using, for instance, the phasor representation of the retarded potential $\tilde{V}$ and the retarded magnetic vector potential $\tilde{A}$ generalized both for a sinusoidally varying volume charge distribution:
\begin{equation}\label{eq08}
\tilde{V}=\frac{1}{4 pi \epsilon} \int_V \frac{\tilde{\rho}}{r}e^{-\jmath \beta r}dV
\end{equation}
and
\begin{equation}\label{eq09}
\tilde{A}=\frac{\mu}{4\pi} \int_V \frac{\tilde{\mathbf{J}}}{r}e^{-\jmath \beta r}dV,
\end{equation}
where $\beta =\omega \sqrt{\mu\epsilon}$ is the wave number of the medium. As is exposed in textbooks (see, e.g., Ref.~\cite{Guru}), from Eqs.~\ref{eq08}-~\ref{eq09} we can obtain the electric and magnetic fields, showing terms that vary as $1/r$ (radiation field), $1/r^2$ (induction terms), and $1/r^3$ (electrostatic field terms). Therefore, as it is shown in Ref.~\cite{Obara_2000}, the propulsive force can result from the near-field and$/$or the far-field (radiative) mode, that is, it is not always the radiative mode of propulsion that can be useful under the point of view of practical engineering.

Recent interest in this new possible mode of field propulsion motivates this paper, mainly the following Refs.~\cite{Charrier,Brady_NASA,Juan,LaFleur}.

\section{Theory}

In Classical Mechanics, contact forces (e.g., gas pressure, fluids, pushing or pulling) obey to the action-to-reaction law and may be regarded as a generalization from familiar experience (action in a continuous medium). They reflect the symmetry of actions when experienced from two different inertial frames and the interaction is instantaneous. However, when electromagnetic forces are actuating due to the action of their representative fields ($\mathbf{E}, \mathbf{B}$), the nature of the problem is another, since these fields sit (``hook") in the physical vacuum. We treat in this Letter the general problem of electromagnetic propulsion by unbalance of forces, DEF (Differential Electromagnetic Forces).

\subsection{The Electromagnetic field and the physical vacuum}

Electromagnetic fields seat on space, on the physical vacuum, that possess a hidden structure still needing to be unveiled in all its complexity, although it is clear that via experimental procedure it is already possible to measure the earth's motion relative to the universe as a whole~\cite{Muller_1978}. Consequences of this hidden structure are studied in the frame of Quantum Electrodynamics: Casimir and Unruh-Davies effects, Zitterbewegung, Lamb shift. Field theories have considerably elucidated the structure of the vacuum, and can be subject to physical experiments, in particular, allowing the creation of real particles in strong (static) external fields, since the normal vacuum state is unstable, decaying into new vacuum state (charged vacuum) containing real particles. Strongly space and$/$or time dependent external fields, above a critical potential strength, the virtual particle pairs generated as fluctuations of the vacuum (allowed by the time-energy uncertainty principle) and with a lifetime of the order of $\Delta t \sim \hbar/m_0c^2$. If this virtual pair is separated during this time $\Delta t$ by a distance superior the Compton wavelength $\lambdabar = \hbar/mc$ (hence, potential and its gradient sufficiently strong $E_{c}$) and if they gain kinetic energy above twice its rest mass $\lambdabar e E_{cr} \sim 2mc^2$, then the pair may become real.

\subsection{DEF: Differential Electromagnetic Forces}

As shown by Harpaz and Soker~\cite{Harpaz1,Harpaz2}, the existence of EM radiation is due to the relative acceleration between the charge and its electric field. The field is induced on space by a charge or electric current and the disturbance creates the EM wave that propagates in the physical vacuum with velocity $c=1/\sqrt{\epsilon_0 \mu_0}$.

Differential electromagnetic forces emerge in out of equilibrium (or better, ``beyond-equilibrium", in the sense that it is the exact purpose where we want to go~\cite{Ottinger_2005}. A physical system characterized by {\it nonlocal} interactions and {\it finite speed of interactions} are beyond-equilibrium; they don't obey to the action-to-reaction law~\cite{Pinheiro1}.

Let us consider two sub-systems (1) and (2) belonging to a closed body, in interaction one with another through the agency of EM forces:
\begin{eqnarray}
  F_{12}(t) &=& P_1(t)P_2 \left( t - \frac{\delta}{c} \right) \\\label{eq1a}
  F_{21}(t) &=& P_1 \left( t - \frac{\delta}{c} \right)P_2(t). \label{eq1b}
\end{eqnarray}
Here, $F_{21}(t)$ is the force exerted on the second body by the first, and vice-versa, $F_{12}(t)$ is the force exerted on the first body by the second, both at a given instant of time $t$.
At first order, their total sum $F(t)=F_{12}(t)+F_{21}(t)$ is given by
\begin{equation}\label{eq2}
F(t)=2P_1(t)P_2(t)-\frac{\delta}{c} \left( \left[\frac{dP_1}{dt} \right]_{\tau=\delta/c} + \left[ \frac{dP_2}{dt} \right]_{\tau=\delta/c}  \right)+ \mathcal{O}(2).
\end{equation}
Using a complex representation for the harmonic response, such as
\begin{eqnarray}
  P_1(t) &=& P_1 e^{j \omega t} \\\label{eq3a}
  P_2(t) &=& P_2 e^{j (\omega t + \alpha)}. \label{eq3b}
\end{eqnarray}
it is obtained (assuming $P_1=P_2$ to simplify), the real force
\begin{widetext}
\begin{equation}\label{eq4}
F(t)=\Re \{\overline{F}(t) \} = P_1^2 \left[ 2\cos (2 \omega t + \alpha) + \frac{\omega \delta}{c} \sin \left( \frac{\omega \delta}{c} \right) + \frac{\omega \delta}{c} \sin \left( \frac{\omega \delta}{c} + \alpha \right) \right],
\end{equation}
\end{widetext}
where $\overline{F}$ means the complex representation of the total force.

We may consider the last two terms of Eq.~\ref{eq4} as the remnants of the retard (field of) force acting from one body over the other, at time $t=\delta/c$, such as:
\begin{eqnarray}\label{eq4a}
F_{12}^1=P_1P_2\frac{\omega \delta}{c}\sin \left( \frac{\omega \delta}{c}  \right) \\
F_{21}^1=P_1P_2\frac{\omega \delta}{c} \sin \left( \frac{\omega \delta}{c}+ \alpha \right).
\end{eqnarray}
Hence, when $\alpha=0$, we have $F_{12}^1=F_{21}^1$, that is, there is an apparent attraction; when there is an opposite phase $\alpha=\pi$, then $F_{12}^1=-F_{21}^1$, which means there is repulsion.

We may notice some similarity with the role played by the electric charge when dealing with electric forces. The above model is completely general, and we have not made any assumption about the form of the functions $P$. We may refer the precursor work of Bjerknes~\cite{Bjerknes_1905} where an account is made of experiments with pulsating or oscillating bodies in a surrounding homogenous fluid.

\begin{figure}
  \includegraphics[width=3.5 in]{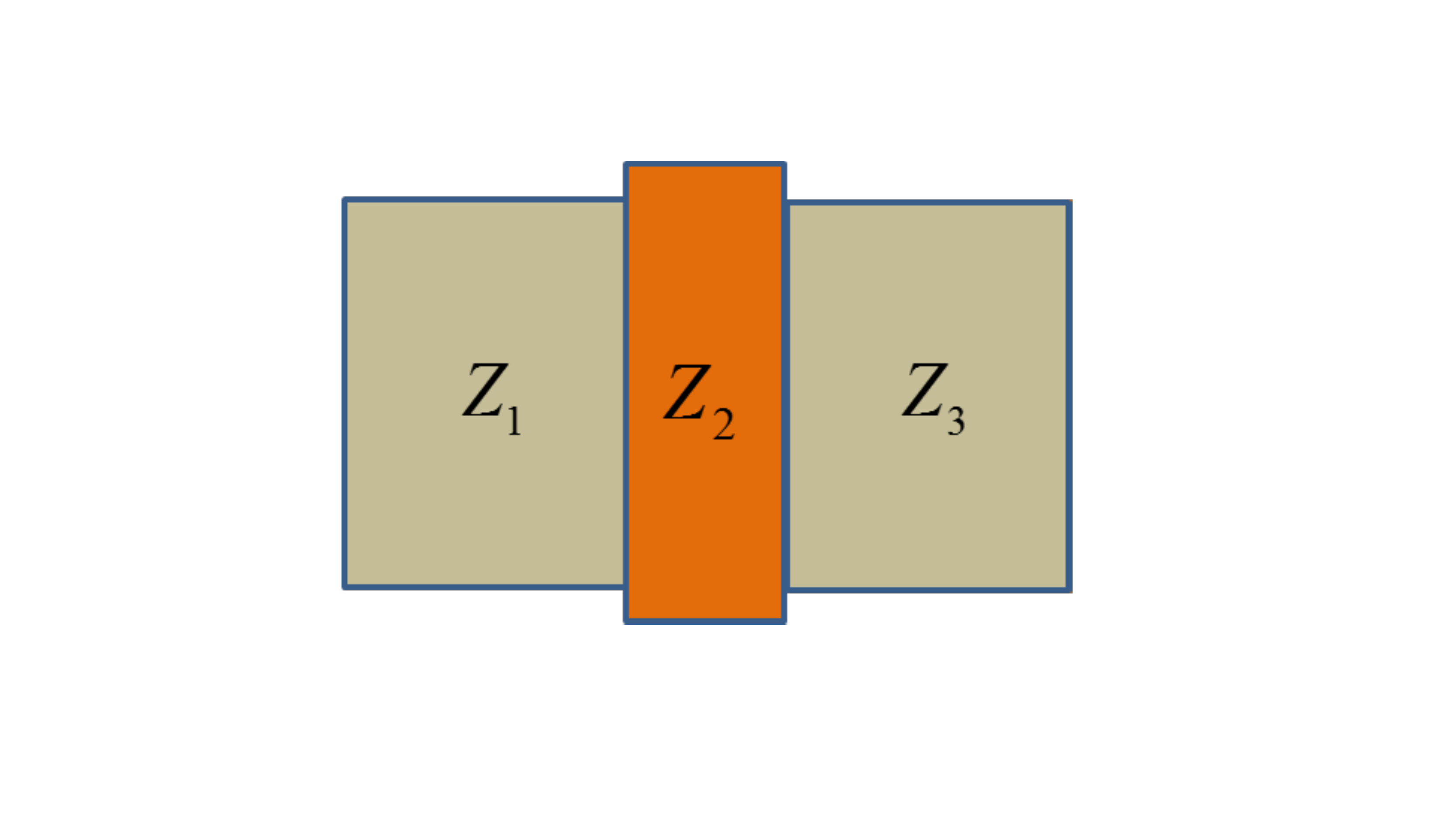}\\
  \caption{Impedance matching in a transmission line.}\label{Fig1}
\end{figure}

Averaging over time, and using a well-known trigonometric identity, it gives
\begin{equation}\label{eq5}
\langle F(t) \rangle = 2 P_1^2 \left( \frac{\omega \delta}{c} \right) \sin \left( \frac{\omega \delta}{c} + \frac{\alpha}{2} \right) \cos \left( \frac{\alpha}{2} \right).
\end{equation}
Evidently Eq.\ref{eq5} is in general non null, and the maximum force is attained when the following conditions are verified:
\begin{eqnarray}
  \alpha &=& 2n_1 \pi - \frac{\pi}{2}; n_1 \in \mathbb{Z} \\
  \frac{\omega \delta}{c}+\frac{\alpha}{2} &=& n_2 \frac{\pi}{2}; n_2 \in \mathbb{Z} \\ \label{BC2}
  \delta &=& \frac{\lambda}{4} (n-\frac{1}{2}); n=n_2-2n_1=1,2,3... \label{BC3}
\end{eqnarray}
We must refer that there are mathematically three different solutions that maximize Eq.~\ref{eq5} but conditions 12 and ~\ref{BC3} have physical meaning.
It is clear from Eq.~\ref{eq5} that the action-to-reaction law is well verified when the interaction between sub-systems are instantaneous ($c \to \infty$):
\begin{equation}\label{eq7}
\langle F(t) \rangle=\langle F_1(t) + F_2(t) \rangle =0 \Rightarrow \langle F_1(t) \rangle = - \langle F_2(t) \rangle.
\end{equation}
DEF has the following properties: i) it is proportional to the square of the intensity $P$; ii) to the angular wave frequency $\omega$; iii) to the space between the sub-systems $\delta$; iv) it is inversely proportional to the speed of interaction; v) when the two sub-systems are out-of-phase, $\alpha= \pm \pi$, and DEF is null, but is maximum when the interacting sub-systems are quadrature.

\subsubsection{The self-accelerated electric dipole: Cornish model}

\begin{figure}
  \includegraphics[width=3.35 in]{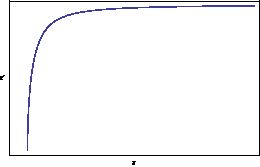}\\
  \caption{Velocity versus position for initial conditions: $\dot{x}(0)=0$, $x(0)=0$.}\label{Fig4}
\end{figure}

\begin{figure}
  \includegraphics[width=4.0 in]{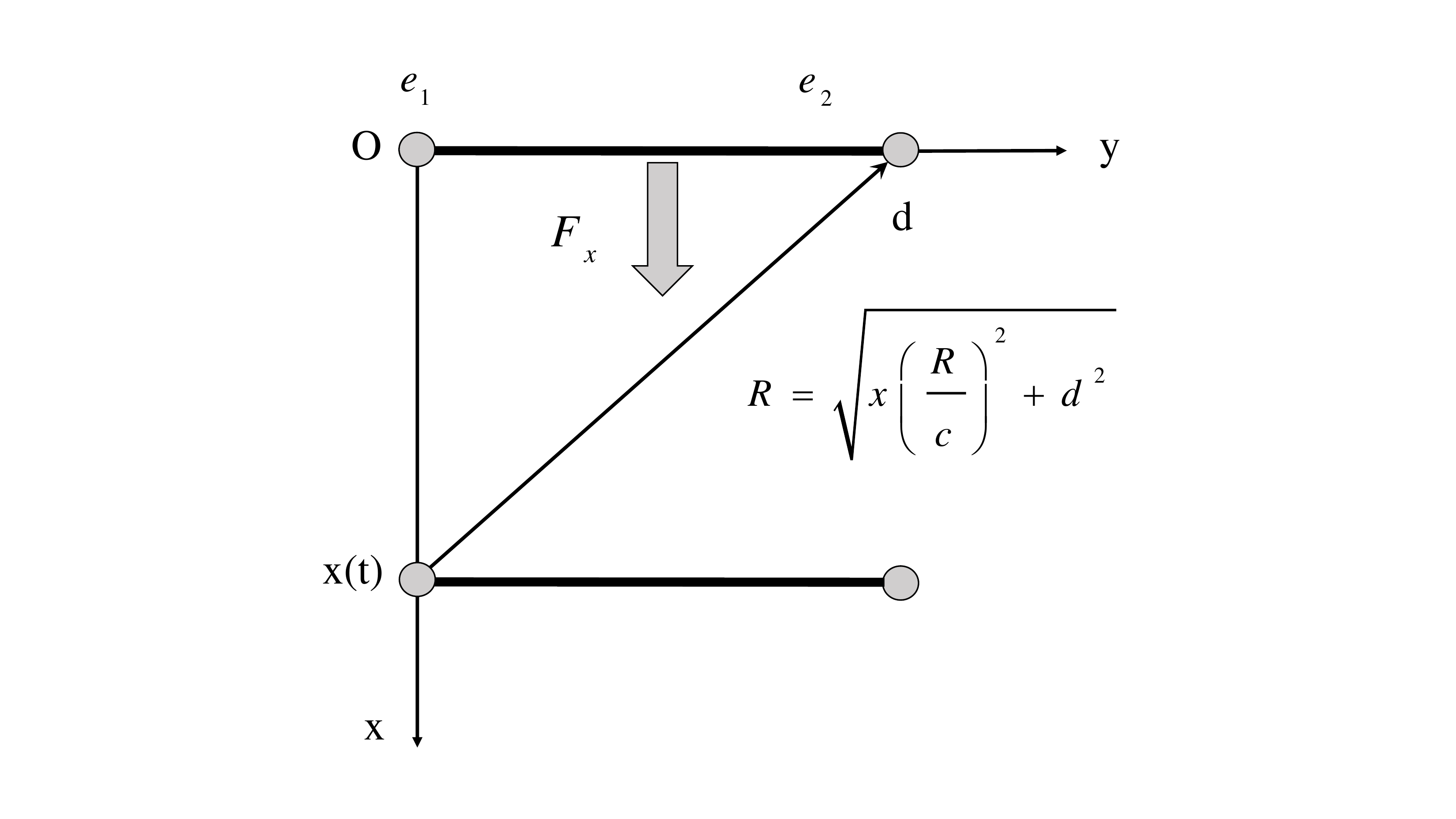}\\
  \caption{The dipole in self-excited motion.}\label{Fig2}
\end{figure}

The motion of simple dumbbell systems formed by two charges $e_1$ and $e_2$ located at a fixed distance $d$ apart and moving in a straight line perpendicular to their connecting line has been studied due to several interesting properties, among them, the possibility of self-excited runaway motion~\cite{Cornish_1985}. It was shown that, in hyperbolic motion, the dipole is accelerated according to the relativistic equation of motion (see Fig~\ref{Fig2}.):
\begin{widetext}
\begin{equation}\label{eq8}
\left( \frac{d}{dt} \left\{ m\dot{x} \left[1 - \left(\frac{\dot{x}}{c} \right)^2 \right]^{-1/2} \right\} \right)_{t=R/c} = \left( \frac{e_1e_2}{R^3} \right) \left[x \left(\frac{R}{c}\right)-\left(\frac{\ddot{x}(0)}{c^2} \right)d^2 \right].
\end{equation}
\end{widetext}
Eq.~\ref{eq8} and Eq.~\ref{eq5} may give terms with similar meaning since both result from developments in terms of the type $t=R/c$. We may start to notice that Eq.~\ref{eq5} is the product of trigonometric functions and assuming $P_1 P_2=\frac{e_1e_2 d}{R^3}$ and considering that $R^3=\frac{e_1e_2d^2}{2mc^2}$ (see Ref.~\cite{Cornish_1985}), then we obtain from Eq.~\ref{eq5}
\begin{equation}\label{eq9}
\langle F \rangle \simeq \frac{2mc^2}{d} \left( 1-\frac{\omega^2 \delta^2}{3! c^2}  \right) + \mathcal{O} \left( \frac{\omega^3 \delta^3}{c^3} \right).
\end{equation}
Notice that $d = \delta$. We may inquiry about which applied frequency can be optimal for DEF propulsion? Eq.~\ref{eq9} suggests a threshold for the applied frequency. We may instead consider propagation along a dielectric medium with index $n=1000$ (e.g., titanate barium strontium):
\begin{equation}\label{eq10}
\omega_c \approx \sqrt{3!}\frac{c}{n \delta} = 1.17 ~\text{MHz},
\end{equation}
otherwise, the applied signal must be of the order of GHz, a frequency domain difficult to attain by actual technology. Fig.~\ref{Fig4} shows the graphic solution of Eq.~\ref{eq8} in the classical limit, assuming $\frac{e_1e_2}{m}=1$, $d=1$ and $\frac{\ddot{x}(0)d^2}{c^2}=10$, in arbitrary units. It is clear that the dipole is initially submitted to a fast acceleration but after a short period of time its velocity becomes nearly uniform.

\begin{figure}
  \includegraphics[width=3.5 in]{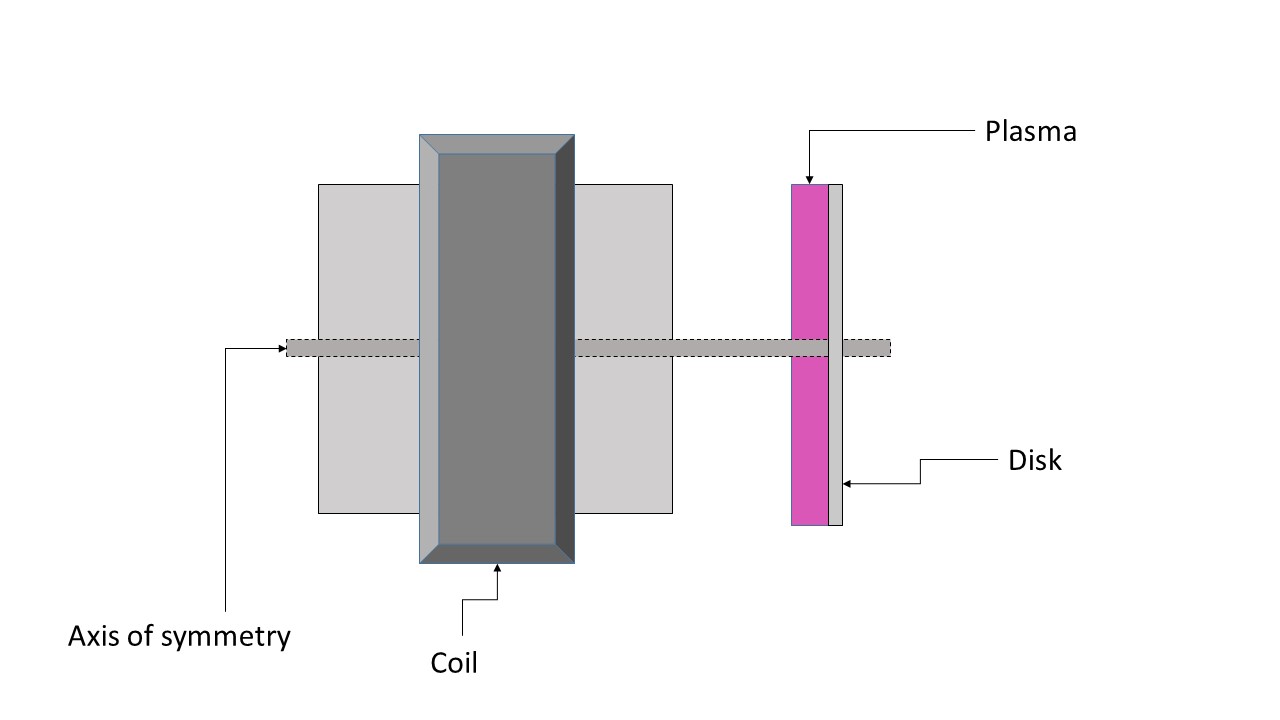}\\
  \caption{Device with two interacting sources (pole), dielectric transmission medium and final plasma module for stochastic acceleration.}\label{Fig3}
\end{figure}

Alternatively, when looking at Eq.~\ref{eq8} we see the x-component of the acceleration. We may envisage a relatively simple process of acceleration, or stochastic acceleration, by generating a plasma on the surface of the second poles of interaction, like shown in Fig.~\ref{Fig3}.

This suggest a possible mechanism to improve the total differential force: generating a plasma by means of RF source and applying an external frequency to stochastically accelerate this sheath of plasma, imposing an RF electric field such as $\ddot{x}(0)=\omega^2 x(0)$.

\subsubsection{Two-electric dipole system: Obara and Baba model}

The same physical principles, as delineated above, are evident in the model of Obara and Baba~\cite{Obara_2000}. The authors discuss the different configurations of two electric dipoles interacting at a distance $d$ of each other and with a phase $\alpha$ between the two: the collinear type array model (CA) and the parallel type array model (PA). The maximum force occurs when the phase difference $\alpha$ of the dipoles is $-\pi/2$ or $3 \pi/2$, in agreement with our above discussion (Eqs.10-~\ref{BC3}). It is shown that the near-field always contribute to propulsion
but only in the case of the parallel type array model the electromagnetic radiation contributes to propulsion. The strongest contribution to the propulsive force comes from the near-field region where the reactive components are very large with respect to the radiating fields.

\subsection{Transmission line and impedance matching}

If we rewrite the total force under the form:
\begin{equation}\label{eq11}
\langle F(t) \rangle =  Zv,
\end{equation}
then it may be associated an impedance of the wave propagation in the medium:
\begin{equation}\label{eq12}
Z=2P_1^2 \left( \frac{k \delta}{c} \right) \sin \left( \frac{\omega \delta}{c} + \frac{\alpha}{2} \right) \cos \left( \frac{\alpha}{2} \right).
\end{equation}
From this viewpoint, impedance matching is a tuning concept to optimize the differential electromagnetic forces. If we have two interacting energy sources characterized by impedances $Z_1$ and $Z_3$, the medium between the two sources have impedance $Z_2$, see Fig.~\ref{Fig1}. Using general concepts of transmission lines, the middle section must be a quarter-wave length section, with impedance $Z_2=\sqrt{Z_1 Z_3}$, in order to transfer the maximum energy from one section to another.

\section{Conclusion}

We gave the general characteristics and properties of a device working on the conversion of electrical energy to kinetic energy in result from an interaction between two sub-systems in an non-instantaneous and nonlocal operative mode. The approach is of interest to practical engineering principles of propulsion since it offers guiding principles useful to build prototypes. The differential electromagnetic force (DEF) depends on the following physical quantities and parameters:

\begin{itemize}
  \item the phase difference $\alpha$ between the two interacting systems (e.g., coil + disk) must be -$\pi/2$ or $3\pi/2$;
  \item the distance between the two systems must be of the order of the wavelength, otherwise fields practically don't interact with each others, because the fields are locally very weak (decreasing like $1/r^2$ for static, $1/r^3$ for static and inductive fields) since it matters to the propulsion effect that the near-fields (supported in the physical vacuum, or recurring to the Maxwell's concepts of field, supported by the Aether), and possibly the far-field radiative EM field (depending on the configuration). However, DEF don't work on the basis of EM radiation, like is supposed to propel the NASA and Chinese projects. IT is supposed to work on the near-fields propelling effect;

  \item The length of the device can be adjusted with the purpose to maximize the DEF effect (and naturally driven by the current technological limitations), to scale the voltage of the signal according to the ratio: $V_1/d_1=V_2/d_2$. Therefore, if we want to match the power with the device, this scaling must be applied , meaning here that we need to increase the voltage signal (and the electric power) accordingly.
\end{itemize}

\begin{acknowledgments}
The author gratefully acknowledge partial financial support by the Funda\c{c}\~{a}o para a Ci\^{e}ncia e Tecnologia under contract ref. SFRH/BSAB/1420/2014. We would like to thank Dr. Manuel Alonso, from IPFN, Lisbon, the suggestion of the designation Differential EM force (DEF).
\end{acknowledgments}

\end{document}